\documentclass[11pt, reqno]{elsarticle}%
\usepackage{lipsum}
\makeatletter
\def\ps@pprintTitle{%
 \let\@oddhead\@empty
 \let\@evenhead\@empty
 \def\@oddfoot{}%
 \let\@evenfoot\@oddfoot}
\makeatother

\usepackage{amsmath}
\usepackage{tikz}
\usetikzlibrary{arrows.meta}

\usepackage{bbm}
\usepackage{array}
\usepackage{amssymb}
\usepackage{amsthm}
\theoremstyle{plain}

\usepackage{dsfont} 

\usepackage[margin=1in]{geometry}
\usepackage[bookmarks=false]{hyperref}
\usepackage{cleveref}
\usepackage{epstopdf}

\def\+{{\oplus}}
\newcommand{\ord}{\operatorname{ord}}

\newtheorem*{theorem*}{Theorem}

\newcounter{comments}

\begin{document}

\author[a]{Venkata Sai Narayana Bavisetty}
\author[b]{Matthew Wheeler}
\author[c]{Julian Vignes}
\author[d]{Matthew Stephen Jones Jr.}
\author[e]{Ruodan Liu}
\author[e]{Sean Campbell}
\author[e]{Claus Kadelka\corref{cor1}}
\ead{ckadelka@iastate.edu}

\cortext[cor1]{Corresponding author}

\address[a]{Department of Integrative Biology and Physiology, University of California, Los Angeles, CA, 90095, United States}
\address[b]{Department of Medicine, University of Florida, Gainesville, FL, 32611, United States}

\address[c]{Watchung Hills Regional High School, Warren, NJ, 07059, United States
}
\address[d]{Ponte Vedra High School, Ponte Vedra Beach, FL, 32081, United States}
\address[e]{Department of Mathematics, Iowa State University, Ames, IA 50011, United States}

\title{Permutation theory governs attractors of critical Boolean networks with connectivity one}


\begin{abstract}
Boolean networks are widely used to model gene regulatory Attractors of Boolean networks model stable gene-expression patterns, yet deriving their properties from network structure remains an open problem.
We solve this problem for critical $K=1$ networks by showing that their feedback loops induce a permutation whose order bounds the average attractor length both above and below by universal constant factors.
This correspondence allows classical results from combinatorics and number theory to be applied directly to Boolean network dynamics.
We find three distinct asymptotic scales for both average and maximum attractor lengths: typical networks scale as $\exp[(1/8+o(1))(\ln N)^2]$, the ensemble means grow as $\exp[N^{1/3+o(1)}]$, and extremal networks attain $\exp[(1+o(1))\sqrt{N\ln N}]$.
Thus, ensemble averages are governed by rare network realizations and are unrepresentative of typical dynamics.
\end{abstract}

\maketitle

\section*{Introduction}
Gene regulatory networks govern fundamental cellular processes such as differentiation, development, and signaling.
Boolean network models, originally introduced by Kauffman~\cite{kauffmanfirst}, are discrete dynamical systems that provide a simplified yet powerful framework for studying these networks. 
A Boolean network comprises a directed graph whose nodes represent genes and whose edges represent regulatory interactions.
Each node takes a binary state (on or off) and is updated in discrete time steps according to an update rule. 
Starting from any initial state, the network eventually reaches a periodic \emph{attractor}: either a fixed point (period one) or a limit cycle (period greater than one). 
The number and lengths (periods) of attractors are key quantities characterizing the long-term dynamics of the network.
Attractors correspond to stable or recurring patterns of gene expression and are commonly interpreted as distinct cell types or phenotypes. 
Despite their simplicity, Boolean networks have successfully reproduced qualitative features of cell cycling, signaling, and differentiation, attracting substantial interest in both biology and physics~\cite{wang2012boolean, davidich2008boolean,deritei2026hhip}.

The most widely studied class of Boolean networks is the Kauffman $NK$ model, in which each of the $N$ nodes receives input from exactly $K$ nodes and is assigned a Boolean update function drawn independently at random; these functions remain fixed as all nodes are updated synchronously in discrete time~\cite{kauffmanfirst}. 
Mathematical understanding of how network structure constrains dynamics differs sharply across connectivity regimes. 
In the high-connectivity limit $K=N$, the network becomes a random map on $2^N$ states.
Thus, the network structure becomes irrelevant~\cite{derrida1987random}, allowing the analytic characterization of many dynamical properties, such as the expected number of attractors, which grows linearly in $N$. 
By contrast, the low-connectivity regime $K\in\{1,2\}$ has proven far more elusive. 
For example, although the number of attractors has been completely characterized for $K=1$ and bounded from below for $K=2$~\cite{flybjergexact,Drosselcriticalk1network,PhysRevE.72.016110,finkexponential,finkinsights,finkconjecturelongpaper,PhysRevLett.90.098701}, it remains poorly understood how the number and lengths of attractors vary with network structure.

Critical low-connectivity networks are of special interest due to the criticality hypothesis~\cite{kauffmanoriginsoforder,kauffmanfirst}.
This posits that biological regulatory networks operate at the phase boundary between ordered and chaotic dynamics, balancing robustness and adaptability~\cite{aldana2007robustness,balleza2008critical,daniels2018criticality,kadelka2024meta}. 
Criticality occurs when perturbations neither grow nor decay on average.
For $K=1$, this is achieved when every node uses a sensitive update function (identity or negation), where every input change alters the output.
These critical $K=1$ models are very tractable: their sparse, loop-dominated structure admits exact analysis while still exhibiting nontrivial dynamics.
Moreover, their relevance extends beyond $K=1$: In critical Kauffman networks with higher connectivity, almost all nodes eventually freeze~\cite{BASTOLLA1998203}, frequently leaving a smaller dynamically active core whose structure is asymptotically dominated by simple loops just like the critical $K=1$ networks~\cite{drosselreview}.
Understanding critical $K=1$ networks can therefore provide insight into the dynamics of more general critical Boolean networks.

Already in 1988, Flyvbjerg showed that the expected number of attractors in critical $K=1$ networks grows at least as $\exp[0.43\sqrt N]$~\cite{flybjergexact}. 
Fink and Sheldon recently derived that the leading term of the expected number of attractors actually scales as $\exp[0.19N]$~\cite{finkexponential}. 
Drossel showed that the expected average attractor length grows faster than any polynomial power and slower than $\exp[0.37m]$ where $m$ is the number of relevant nodes in feedback loops~\cite{Drosselcriticalk1network}. Sheldon and Fink derived an improved upper bound of $\exp[\sqrt{m}]$~\cite{finkinsights}. While exact formulae for attractor statistics are known for networks with given loop lengths~\cite{finkconjecturelongpaper}, the asymptotic scaling of the average attractor length with network size $N$ remains unresolved.
Additionally, despite the apparent simplicity of critical $K=1$ networks and decades of study~\cite{flybjergexact,drosselreview,Drosselcriticalk1network,loopsjapanesegroup,finkexponential,finkinsights,demongeot2010number}, existing results do not reveal a simple relationship between feedback-loop structure and long-term dynamics. 
Consequently, even for networks with prescribed loop lengths, extracting their attractor statistics remains complicated.

In this article, we show that the feedback-loop structure of a critical $K=1$ network can be encoded as a  permutation $\sigma$, whose order $\ord(\sigma)$
governs the network's long-term dynamics. Inspired by Fink and Sheldon's recent advances~\cite{finkexponential,finkinsights,finkconjecturelongpaper}, we derive that, up to universal constant factors, the maximum and
average attractor lengths are given by $\mathrm{ord}(\sigma)$.
This establishes a direct link between network structure and dynamics and allows classical results from combinatorics and number theory to apply directly to critical $K=1$ networks. 
For instance, applying results of Erd\H{o}s--Tur\'an, Landau, and Goh--Schmutz, we obtain sharply separated scales of long-term dynamics: 
Asymptotically almost surely,  the longest attractor of a random network has length $\exp[(1/8+o(1))(\ln N)^2]$, whereas the longest attainable attractor across all such networks grows as $\exp[(1+o(1))\sqrt{N\ln N}]$.
Meanwhile, both the expected average and expected maximum attractor lengths grow as $\exp[N^{1/3+o(1)}]$.
Thus, typical, ensemble-averaged, and extremal behavior occur on parametrically different scales. 
In particular, the ensemble mean is governed by rare network realizations that are unlikely to be captured by standard random sampling, rendering naive Monte Carlo estimation of mean dynamical quantities unreliable.
\section*{Permutation encodes feedback-loop structure}
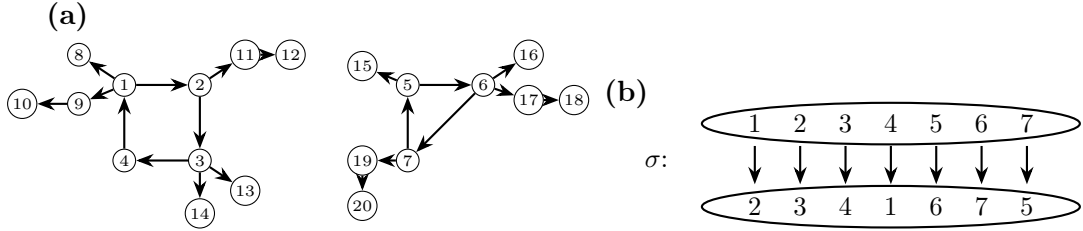
\begin{figure}[h]
\centering
\begin{tikzpicture}[scale=0.5,
    every node/.style={circle, draw, minimum size=0.3cm, font=\tiny, inner sep=1pt},
    arrow/.style={-{Stealth}, thick},
    node distance=2.0cm
]
\node (1) at (0,0)   {$1$};
\node (2) at (2,0)   {$2$};
\node (3) at (2,-2)  {$3$};
\node (4) at (0,-2)  {$4$};
\draw[arrow] (1) -- (2);
\draw[arrow] (2) -- (3);
\draw[arrow] (3) -- (4);
\draw[arrow] (4) -- (1);
\node (T1a) at (-1.2,  0.8) {$8$};
\node (T1b) at (-1.2, -0.5) {$9$};
\node (T1c) at (-2.7, -0.5) {$10$};
\draw[arrow] (1) -- (T1a);
\draw[arrow] (1) -- (T1b);
\draw[arrow] (T1b) -- (T1c);
\node (T2a) at (3.2,  0.8) {$11$};
\node (T2b) at (4.4,  0.8) {$12$};
\draw[arrow] (2) -- (T2a);
\draw[arrow] (T2a) -- (T2b);
\node (T3a) at (3.2, -2.8) {$13$};
\node (T3b) at (2.0, -3.4) {$14$};
\draw[arrow] (3) -- (T3a);
\draw[arrow] (3) -- (T3b);
\begin{scope}[xshift=7.5cm]
\node (5) at (0,0)   {$5$};
\node (6) at (2,0)   {$6$};
\node (8) at (0,-2)  {$7$};
\draw[arrow] (5) -- (6);
\draw[arrow] (6) -- (8);
\draw[arrow] (8) -- (5);
\node (S1a) at (-1.2,  0.5) {$15$};
\draw[arrow] (5) -- (S1a);
\node (S2a) at (3.2,  0.8)  {$16$};
\node (S2b) at (3.2, -0.4)  {$17$};
\node (S2c) at (4.4, -0.4)  {$18$};
\draw[arrow] (6) -- (S2a);
\draw[arrow] (6) -- (S2b);
\draw[arrow] (S2b) -- (S2c);
\node (S3a) at (-1.2, -2.0) {$19$};
\node (S3b) at (-1.2, -3.2) {$20$};
\draw[arrow] (8) -- (S3a);
\draw[arrow] (S3a) -- (S3b);
\end{scope}
\node[draw=none, font=\normalsize\bfseries] at (-1.5, 1.8) {(a)};
\end{tikzpicture}
\vspace{1.0em}
\begin{tikzpicture}[
    scale=0.5,
    pnode/.style={draw=none, font=\small, inner sep=2pt},
    parrow/.style={-{Stealth}, thick, shorten >=3pt, shorten <=3pt}
]
\node[pnode, font=\normalsize\bfseries] at (-7.0, 0.8) {(b)};
\draw[thick] (0,0) ellipse (5.0cm and 0.55cm);
\draw[thick] (0,-2.2) ellipse (5.0cm and 0.55cm);
\node[pnode] at (-6.2, -1)    {$\sigma$:};
\foreach \i/\x in {1/-3.6, 2/-2.4, 3/-1.2, 4/0.0, 5/1.2, 6/2.4, 7/3.6} {
    \node[pnode] (top\i) at (\x, 0) {$\i$};
}
\foreach \i/\x/\lbl in {1/-3.6/2, 2/-2.4/3, 3/-1.2/4, 4/0.0/1, 5/1.2/6, 6/2.4/7, 7/3.6/5} {
    \node[pnode] (bot\i) at (\x, -2.2) {$\lbl$};
}
\foreach \i in {1,2,3,4,5,6,7} {
    \draw[parrow] (top\i) -- (bot\i);
}
\end{tikzpicture}
\caption{(a)~Two disjoint $K=1$ Boolean network components, each consisting of
a loop of relevant nodes (numbered $1$--$4$ and $5$--$7$) with trees of
non-relevant nodes branching out. (b)~The corresponding permutation $\sigma$
of the relevant nodes, consisting of loops $(1\;2\;3\;4)$ and $(5\;6\;7)$;
the arrows indicate the image of each node under $\sigma$.}
\label{fig:k1networks}
\end{figure}

The underlying graph of a critical $K=1$ network decomposes naturally into directed feedback loops with trees rooted at the loop nodes (Fig.~\ref{fig:k1networks}a). 
Each edge corresponds to either an identity (activating) or negation (inhibitory) update, and a loop is \emph{positive} if it contains an even number of inhibitory edges and \emph{negative} otherwise. 
The long-term dynamics are governed entirely by the loop nodes, called the \emph{relevant nodes}, while nodes on the trees affect only transient behavior~\cite{flybjergexact,drosselreview}. 
In particular, the number and lengths of attractors depend only on the lengths and signs of the loops~\cite{demongeot2010number,Golomb2017}.

Let $m$ denote the number of relevant nodes. 
Within the subnetwork induced by these nodes, every node has exactly one incoming and one outgoing edge, so its directed edges define a permutation $\sigma\in S_m$. 
For example, the relevant subnetwork in Fig.~\ref{fig:k1networks}a consists of a $4$-loop and a $3$-loop, corresponding to the permutation $(1\;2\;3\;4)(5\;6\;7)\in S_7$~(Fig.~\ref{fig:k1networks}b). 
The permutation cycles correspond exactly to the feedback loops, and the order of the permutation is their least common multiple: $\mathrm{ord}(\sigma)=\mathrm{lcm}(4,3)=12$.

\section*{Permutation order determines long-term dynamics}
Let $F$ be a
critical $K=1$ network whose relevant nodes form loops of lengths
$m_1,\ldots,m_k$, with $m=m_1+\cdots+m_k$ relevant nodes in total. The
restriction $F_{\mathrm{rel}}$ to these $m$ nodes is a bijection on the
$2^m$ relevant-node configurations, so every configuration lies on an
attractor. Because the long-term dynamics depends only on the relevant
nodes, the number of attractors $C(F)$ and the average attractor length
$\bar A(F)$ of $F$ coincide with those of $F_{\mathrm{rel}}$. Since the
$2^m$ attractor states partition into $C(F)$ cycles of average length
$\bar A(F)$, we have
\begin{equation}\label{eq:expected_length}
    \bar A(F)\; C(F)=2^m.
\end{equation}

Exact formulae for these quantities are known~\cite{finkconjecturelongpaper}, but how they are connected to the feedback loop structure has not been demonstrated.
We show that $\bar A(F)$ is determined, up to universal constant factors, by the order of the induced permutation $\sigma$.
By~\cite{flybjergexact}, the maximum attractor length $\ell_{\max}(F)$ satisfies
\begin{equation}\label{eq:maxattractororder}
    \ell_{\max}(F)\in\{\ord(\sigma),\, 2\ord(\sigma)\},
\end{equation}
yielding a trivial upper bound
\begin{equation}
    \bar{A}(F)\leq 2\ord(\sigma).
\end{equation}

For the lower bound, note that every attractor length divides
$2\ord(\sigma)$~\cite{finkconjecturelongpaper}, so Burnside's lemma gives
\begin{equation}\label{eq:numberofattractorsviaburnside}
    C(F)=\frac{1}{2\ord(\sigma)}
    \sum_{t=1}^{2\ord(\sigma)} |\mathrm{Fix}(F_{\mathrm{rel}}^t)|,
\end{equation}
where $\mathrm{Fix}(F_{\mathrm{rel}}^t)$ denotes the set of
relevant states $\mathbf x\in\{0,1\}^m$ satisfying
$F_{\mathrm{rel}}^t(\mathbf x)=\mathbf x$. Because the $k$ loops evolve
independently, the fixed-point count factorizes as
\begin{equation}
|\mathrm{Fix}(F_{\mathrm{rel}}^t)|
=\prod_{i=1}^{k}\left\lvert\operatorname{Fix}(F_{i}^t)\right\rvert,
\end{equation}
where $F_i$ is the restriction to the $i$-th loop.
By~\cite{samuelsson2005random}, for each loop we have
\begin{equation}\label{eq:fixedpointsindividualloops}
    \left\lvert\operatorname{Fix}(F_{i}^t)\right\rvert=
\begin{cases}
2^{\gcd(t,\,m_i)}, & \begin{subarray}{l}
  \text{if the loop is positive}\\
  \text{or } t/\gcd(t,\,m_i) \text{ is even}
\end{subarray}\\[12pt]
0, & \text{otherwise}.
\end{cases}
\end{equation}

In particular, the fixed-point count of a negative loop is
at most that of a positive loop of the same length. That is, for every $t$,
$|\mathrm{Fix}(F_{\mathrm{rel}}^t)|\leq|\mathrm{Fix}(F_{+,\mathrm{rel}}^t)|$,
where $F_+$ is a network obtained from $F$ by making every feedback loop positive while preserving its length. It follows that
$C(F)\le C(F_+)$ and, by Eq.~\eqref{eq:expected_length},
\begin{equation}\label{eq:positive_network_lower_bound}
    \bar A(F)\geq\bar A(F_+).
\end{equation}
In other words, among all networks with a given loop structure, those
with only positive feedback loops have the largest number of attractors
and the smallest average attractor length. It therefore suffices to
establish the lower bound for networks containing only positive
loops. For such a network, every attractor length divides $\ord(\sigma)$, and Eqs.~\eqref{eq:expected_length}, \eqref{eq:numberofattractorsviaburnside},  \eqref{eq:fixedpointsindividualloops} yield
\begin{equation}\label{eq:stop_main_proof_here}
    \bar A(F_+)
    =\ord(\sigma)\cdot \frac{2^{\sum_{i=1}^k m_i}}{\displaystyle
    \sum_{t=1}^{\ord(\sigma)}
    \prod_{i=1}^k 2^{\gcd(t,\,m_i)}}.
\end{equation}

The factor multiplying $\ord(\sigma)$ is bounded below, uniformly over all choices of loop lengths, by the universal constant $R_{\min} \approx 0.32089$ (see Appendix). Thus,
\begin{equation}\label{eq:meanaveragelength}
    R_{\min}\,\ord(\sigma)
    \leq \bar A(F)
    \leq 2\,\ord(\sigma).
\end{equation}
Both constants are optimal: a single negative loop with a length that is a power of two attains the upper bound, while the lower bound $R_{\min}\,\text{ord}(\sigma)$ is approached arbitrarily closely by networks with only positive loops whose lengths are the distinct primes $p \leq x$ as $x\to\infty$.
Equivalently,
\begin{equation}\label{eq:lnmeanaveragelength}
    \ln \bar A(F)=\ln\ord(\sigma)+O(1).
\end{equation}
Together with Eq.~\eqref{eq:maxattractororder}, this implies
\begin{equation}
\label{eq:mean_equal_max_attractor_length}
    \bar A(F)=\Theta\!\left(\ell_{\max}(F)\right).
\end{equation}
Thus, even though most networks possess attractors of vastly different lengths, the average and the maximum attractor lengths have the same asymptotic scaling and are determined up to a universal constant factor by the single structural quantity $\ord(\sigma)$.

\section*{Improved scaling law for the expected number of attractors}
Let $\mathbb E_N[C]$ denote the expected number of attractors over the full ensemble of $N$-node critical $K=1$ networks.
Fink and Sheldon~\cite{finkexponential} showed that
\begin{equation}
    \ln \mathbb{E}_N[C]
    =N\ln\!\left(2/\sqrt e\right)+o(N).
\end{equation}
The aforementioned connection to permutation order allows us to sharpen the additive error from $o(N)$ to $O(\ln N)$.
To see this, observe that in a network $F$ with $m$ relevant nodes,  Eqs.~\eqref{eq:expected_length} and~\eqref{eq:meanaveragelength} yield 
\begin{equation}
    {2^{m-1}}/{\ord(\sigma)}
    \leq C(F) \leq 2^m.
    \label{eq:count-bound}
\end{equation}
Conditional on having $m$ relevant nodes, the induced permutation is uniformly
distributed over $S_m$~\cite{samuelsson2005random}.
Among these $m!$ permutations, $(m-1)!$ permutations are cyclic with order $m$, so that 
$\mathbb{E}_m[1/\ord(\sigma)]\geq 1/m^2$.
Thus, averaging Eq.~\eqref{eq:count-bound} over all networks with $m$ relevant nodes gives
    $2^{m-1}/m^2
    \leq \mathbb{E}_m[C]
    \leq 2^m$.
Averaging over the distribution 
\begin{equation}\label{eq:p_N_m}
P_N(m) = \left(\frac{m}{N}\right) \left(1 - \frac{1}{N}\right) \dots \left(1 - \frac{m-1}{N}\right)
\end{equation}
of the number of relevant nodes in an $N$-node critical $K=1$ network~\cite{flybjergexact} then yields
\begin{equation}\label{eq:endmatter_startb}
    \sum_m\frac{P_N(m)}{m^2}\,2^{m-1}
    \leq \mathbb{E}_N[C]
    \leq \sum_m P_N(m)\,2^m.
\end{equation}
Both bounds have logarithms equal to
$N\ln(2/\sqrt e)+O(\ln N)$ (see Appendix). Thus, we get
\begin{equation}\label{eq:expectednumber}
    \ln \mathbb{E}_N[C]
    =N\ln\!\left({2}/{\sqrt e}\right)+O(\ln N).
\end{equation}
This sharpens the previously known $o(N)$ correction to 
$O(\ln N)$, restricting deviations from the leading exponential 
growth to polynomial factors in $N$.

\section*{Scaling of expected average and maximum attractor lengths}
For a uniformly random permutation on $m$ elements, Goh and Schmutz~\cite{schmutz} proved that
\begin{equation}
    \ln \mathbb{E}_m[\ord(\sigma)]
    =c\sqrt{{m}/{\ln m}}\,(1+o(1)),
\end{equation}
and Stong computed $c\approx2.99047$~\cite{stong1998average}.
Since Eq.~\eqref{eq:meanaveragelength} bounds
$\bar A(F)/\ord(\sigma)$ by universal constants, we get
\begin{equation}\label{eq:EmAbar}
    \ln \mathbb{E}_m[\bar A]
    =
    c\sqrt{{m}/{\ln m}}\,(1+o(1))
    =
    m^{1/2+o(1)}.
\end{equation}
Averaging over $P_N(m)$ (Eq.~\eqref{eq:p_N_m}) and using a saddle-point argument (see Appendix) gives
\begin{equation}\label{eq:full_asymptotic}
    \mathbb{E}_N[\bar A]
    =\exp\!\left[N^{1/3+o(1)}\right].
\end{equation}
Moreover, Eq.~\eqref{eq:mean_equal_max_attractor_length} bounds
$\bar A(F)$ and $\ell_{\max}(F)$ by universal constant factors for
every network, so their ensemble means have the same asymptotic scaling
\begin{equation}\label{eq:full_asymptotic_maximum}
    \mathbb{E}_N[\ell_{\max}]
    =\exp\left[N^{1/3+o(1)}\right].
\end{equation}

\section*{Typical, average, and extremal attractor lengths}

Erd\H{o}s and Tur\'an~\cite{ErdosTuran1965} established that the order of a uniformly random permutation $\sigma\in S_m$ has the asymptotic behavior
\begin{equation}
\ln\ord(\sigma)
=
\left(1/2+o(1)\right)(\ln m)^2
\end{equation}
in probability as $m\to\infty$. Meanwhile, Eq.~\eqref{eq:p_N_m} yields
\begin{equation}
\ln m
=
\left(1/2+o(1)\right)\ln N
\end{equation}
in probability as $N\to\infty$.
Since
$\ell_{\max}(F)$ is $\ord(\sigma)$ or $2\ord(\sigma)$, it follows that
\begin{equation}\label{eq:typical_maximum}
\ell_{\max}
=
\exp\!\left[
\left(1/8+o(1)\right)(\ln N)^2
\right]
\end{equation}
in probability as $N\to\infty$. Thus, the typical maximum attractor length is sub-exponential in $N$ with leading asymptotic $\exp[(\ln N)^2/8]$.

\begin{figure}
    \centering
    \includegraphics[width=\linewidth]{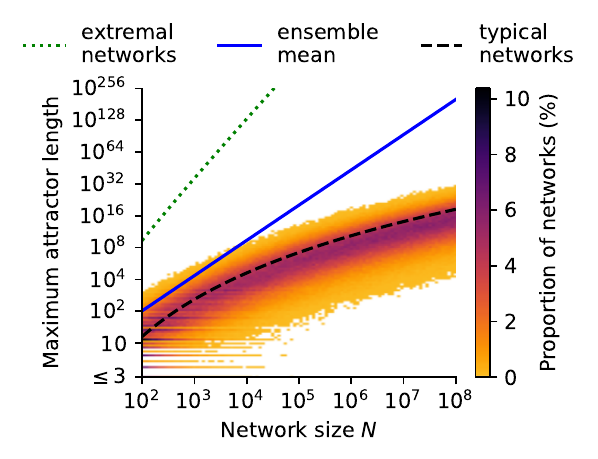}
    \caption{Distribution of maximum attractor lengths $\ell_{\max}$ in critical $K=1$ networks as a function of network size $N$. Color indicates the percentage of $10^4$ independently sampled networks at each of 100 logarithmically spaced values of $N$. Both axes are logarithmic, with the attractor-length axis additionally logarithmically compressed to accommodate the broad range of scales. Curves show the leading asymptotic scales of maximum attractor length for typical networks (black dashed, Eq.~\ref{eq:typical_maximum}), the ensemble mean (blue solid, Eq.~\ref{eq:full_asymptotic_maximum}), and extremal networks (green dotted, Eq.~\ref{eq:max_max_attr_landau}); the average attractor length has the same asymptotic scales by Eq.~\eqref{eq:mean_equal_max_attractor_length}.
    }
    \label{fig:discrepancy_new}
\end{figure}

At the opposite extreme, Landau~\cite{Landau1903} showed that the largest possible order of a permutation of $m$ elements satisfies
\begin{equation}\label{eq:landau}
\max_{\sigma\in S_m}\ln\ord(\sigma)
\sim \sqrt{m\ln m},
\end{equation}
a result previously connected to critical $K=1$ networks by Fink~\cite{finkconjecturelongpaper}.
Since any permutation in $S_N$ can be realized by the feedback loops
of an $N$-node critical $K=1$ network with all nodes relevant, by Eqs.~\eqref{eq:maxattractororder} and \eqref{eq:landau} the
longest attractor length among all such networks is given by
\begin{equation}\label{eq:max_max_attr_landau}
\max_F \ell_{\max}(F)
=
\exp\left[(1+o(1))\sqrt{N\ln N}\right].
\end{equation}
Together with Eq.~\eqref{eq:full_asymptotic_maximum}, these results reveal three fundamentally different asymptotic scales of maximum attractor length corresponding to typical, ensemble-averaged, and extremal behavior (Fig.~\ref{fig:discrepancy_new}).
By Eq.~\eqref{eq:mean_equal_max_attractor_length}, the average attractor length exhibits the same three asymptotic scales.
This means that ensemble averages are governed by rare network realizations with exceptionally long attractors and are therefore unrepresentative of typical network dynamics.

\section*{Discussion}
We have shown that the long-term dynamics of critical $K=1$ Boolean networks are governed, up to universal constant factors, by a single structural quantity: the order of the permutation induced by the network's feedback loops. Previous work established scaling laws for the expected number of attractors and asymptotic upper bounds for the expected average attractor length~\cite{finkexponential,finkinsights}. 
Our results identify the network-level structural quantity controlling these attractor statistics, enabling detailed insights into typical, ensemble-averaged, and extremal behavior.
The resulting separation is striking: almost all networks have average and maximum attractor lengths on the scale $\exp[(1/8+o(1))(\ln N)^2]$, whereas the ensemble mean grows as $\exp[N^{1/3+o(1)}]$, and extremal networks attain $\exp[(1+o(1))\sqrt{N\ln N}]$.
Thus, the ensemble mean describes a qualitatively different part of the network ensemble than a typical realization.

This distinction has important consequences for computational studies of Boolean network dynamics.
Because the ensemble mean is governed by rare networks with exceptionally large permutation order, standard random sampling is unlikely to encounter the realizations that contribute disproportionately to the expectation.
Naive Monte Carlo simulations can therefore substantially underestimate average and maximum attractor lengths. The natural next steps would be to characterize the full distribution of attractor lengths and develop sampling methods that efficiently resolve its rare-event tail.
Crucially, central limit theorems for permutation orders~\cite{ErdosTuran1967a,ErdosTuran1967b,ErdosTuran1968, zacharovas2004distribution,manstavivcius2011limit,ford2021cycle}, provide a foundation for this task, offering insights into the broader distribution of attractor statistics that go beyond what the expectation reveals.

More generally, our permutation-theoretic viewpoint may extend beyond the $K=1$ setting considered here. In critical $K=2$ Boolean networks, the dynamically relevant core reduces asymptotically to an effective $K=1$ structure~\cite{drosselreview}, suggesting that permutation-based methods may provide a route toward resolving the long-standing problem of attractor statistics in critical Boolean networks with higher connectivity.

\textit{Acknowledgements.}---The authors thank Maria Siskaki and Ahana Ghosh for useful comments and suggestions. 
This research was supported in part by grants from the NSF (DMS-2235451) and Simons Foundation (MPS-NITMB-00005320) to the NSF-Simons National Institute for Theory and Mathematics in Biology (NITMB). R.L., S.C., and C.K. were further supported by the NSF (DMS-2424632 and DMS-2451973). M.W. was further supported by the NSF (DMS-2424635) and the NIH (R01-AI135128 and R01-HL169974-01). 

\textit{Data availability.}---The code underlying Figure~2 is openly available on Zenodo~\cite{kadelka2026code}.

\section*{Appendix}
\setcounter{equation}{0}
\renewcommand{\theequation}{A\arabic{equation}}

Here we provide proofs of the principal results stated in the article.
Throughout, let $m_1,\ldots,m_k$ denote the lengths of the feedback
loops, and let $m=\sum_i m_i$ denote the number of relevant nodes.

\subsection*{Universal lower bound}
Let $
L:=\operatorname{lcm}(m_1,\ldots,m_k)=\operatorname{ord}(\sigma).$ It remains to show that the factor
\begin{equation}
\mathcal A(m_1,\ldots,m_k)
:=
\frac{2^{\sum_{i=1}^k m_i}}
{\displaystyle\sum_{t=1}^{L}
\prod_{i=1}^k2^{\gcd(t,m_i)}},
\label{eq:A_end}
\end{equation}
appearing in Eq.~\eqref{eq:stop_main_proof_here}, is bounded below by a universal positive constant. To prove this, we first derive four number-theoretic properties of $\mathcal A$.

\noindent First, if $m_1=ab$ with $\gcd(a,b)=1$, then
\begin{equation}
\mathcal A(m_1,m_2,\ldots,m_k)
\geq
\mathcal A(a,b,m_2,\ldots,m_k),
\label{eq:split_end}
\end{equation}
which follows from
\begin{equation}
\gcd(t,ab)
\leq
\gcd(t,a)+\gcd(t,b)+ab-a-b.
\end{equation}

\noindent Second, if
$m_1$ divides $\operatorname{lcm}(m_2,\ldots,m_k)$, then
\begin{equation}
\mathcal A(m_1,m_2,\ldots,m_k)
\geq
\mathcal A(m_2,\ldots,m_k),
\label{eq:delete_end}
\end{equation}
since $\gcd(t,m_1)\leq m_1$ termwise in Eq.~\eqref{eq:A_end}.

\noindent Third, for distinct primes $p_1,\ldots,p_n$, we have 
    \begin{equation}\label{eq:nt_4}
    \mathcal{A}(p_1^{a_1},\ldots,p_n^{a_n})=\prod_{i=1}^n\mathcal{A}(p_i^{a_i}),
    \end{equation}
    since
    \begin{equation}
    \sum_{j=1}^L\prod_i 2^{\gcd(j,p_i^{a_i})}=\prod_i\sum_{j=1}^{p_i^{a_i}}2^{\gcd(j,p_i^{a_i})},
    \end{equation}
 which follows from the Chinese remainder theorem.
 
\noindent Fourth, for any prime $p$ and integer $a\geq1$,
\begin{equation}\label{eq:nt_3}
\mathcal A(p^a)\geq\mathcal A(p).
\end{equation}
Indeed,
\begin{equation}
\mathcal A(p^a)
=
\frac{2^{p^a}}
{2^{p^a}+
 \displaystyle\sum_{j=1}^{a}
 p^{j-1}(p-1)2^{p^{a-j}}}.
\end{equation}
After cross-multiplication and cancellation of $p-1$, Eq.~\eqref{eq:nt_3} is equivalent to
\begin{equation}\label{eq:nt_3_equiv}
2^p\sum_{j=1}^{a}p^{j-1}2^{p^{a-j}}
\leq 2^{p^a+1}.
\end{equation}
This holds with equality for $a=1$. By induction, if Eq.~\eqref{eq:nt_3_equiv} holds for $a$, then
\begin{equation}
\begin{split}
2^p\sum_{j=1}^{a+1}p^{j-1}2^{p^{a+1-j}}
&=
2^{p^{a}+p}
+
p\,2^p\sum_{j=1}^{a}p^{j-1}2^{p^{a-j}}\\
&\leq
2^{p^a+p}+p\,2^{p^a+1}
=
2^{p^a}(2^p+2p).
\end{split}
\end{equation}
It therefore suffices to show
\begin{equation}
2^{p^{a+1}-p^a}\geq 2^{p-1}+p,
\end{equation}
which follows from
\begin{equation}
2^{p^{a+1}-p^a}
=
2^{p^a(p-1)}
\geq 2^p
\geq 2^{p-1}+p.
\end{equation}
Thus Eq.~\eqref{eq:nt_3_equiv}, and hence Eq.~\eqref{eq:nt_3}, holds for all $a\geq1$.

Let $p_1^{a_1},\ldots,p_n^{a_n}$ be the maximal prime-power divisors of $L=\text{lcm}(m_1,\ldots,m_k)$. Repeated use of Eqs.~\eqref{eq:split_end} and \eqref{eq:delete_end} and subsequent exploitation of the other properties of $\mathcal A$ then gives 
\begin{align}
\mathcal A(m_1,\ldots,m_k)
&\overset{\eqref{eq:split_end},\eqref{eq:delete_end}}{\geq}
\mathcal A(p_1^{a_1},\ldots,p_n^{a_n})
\nonumber\\
&\overset{\eqref{eq:nt_4}}{=}
\prod_{i=1}^n\mathcal A(p_i^{a_i})
\nonumber\\
&\overset{\eqref{eq:nt_3}}{\geq}
\prod_{i=1}^n\mathcal A(p_i)
\nonumber\\
&>
\prod_{p\;\mathrm{prime}}
\frac{2^p}{2^p+2(p-1)}
\nonumber\\
&=
\prod_{p\;\mathrm{prime}}
\frac{1}{1+2(p-1)/2^p}
=:R_{\min},
\label{eq:Rmin_end}
\end{align}
which converges because 
\begin{equation}
\sum_{p\;\text{prime}} 2(p-1)/2^p < \infty.
\end{equation}
Numerically,  $R_{\min}\approx 0.32089$. Moreover, the strict inequality in Eq.~\eqref{eq:Rmin_end} follows because the preceding product contains only the finitely many primes dividing $L$, whereas every omitted factor in the infinite product is strictly less than $1$.
Hence, by Eq.~\eqref{eq:positive_network_lower_bound},
$\bar A(F)\geq \bar A(F_+)\geq R_{\min}L$,
which completes the proof of Eq.~\eqref{eq:meanaveragelength}.
\newline

\setcounter{equation}{0}
\renewcommand{\theequation}{B\arabic{equation}}

\subsection*{Expected number of attractors}
We derive the asymptotics of the two bounds in Eq.~\eqref{eq:endmatter_startb}. Define
\begin{equation}
T_N(m):=
2^m\frac{N!}{(N-m)!\,N^m},
\end{equation}
so that the summands appearing in the lower and upper bound can be expressed as
\begin{equation}
\frac{P_N(m)}{m^2}2^{m-1}=\frac{T_N(m)}{2Nm},
\qquad
P_N(m)2^m=\frac{m}{N}T_N(m).
\end{equation}
For fixed $N$,
\begin{equation}
\frac{T_N(m+1)}{T_N(m)}
=
\frac{2(N-m)}{N},
\end{equation}
so $m_*=\lceil N/2\rceil$ maximizes $T_N(m)$. Retaining only the $m_*$ term gives lower bounds on both sums, while $T_N(m)\leq T_N(m_*)$ for every $m$ gives the corresponding upper
bounds:
\begin{equation}
\frac{T_N(m_*)}{2N^2}
\leq
\sum_m\frac{P_N(m)}{m^2}2^{m-1}
\leq
\frac{T_N(m_*)}{2},
\end{equation}
and
\begin{equation}
\frac 12 T_N(m_*)\leq
\sum_mP_N(m)2^m
\leq
N\,T_N(m_*).
\end{equation}
Thus, both bounds in Eq.~\eqref{eq:endmatter_startb} differ from
$T_N(m_*)$ by at most polynomial factors in $N$, so their logarithms
differ from $\ln T_N(m_*)$ by at most $O(\ln N)$. Stirling's formula
gives
\begin{equation}
\ln T_N(m_*)
=
N\ln\left({2}/{\sqrt e}\right)+O(\ln N),
\end{equation}
which gives the asymptotics of both bounds in Eq.~\eqref{eq:endmatter_startb} and hence proves Eq.~\eqref{eq:expectednumber}.
\newline

\setcounter{equation}{0}
\renewcommand{\theequation}{C\arabic{equation}}

\subsection*{Expected average attractor length} 
We now derive Eq.~\eqref{eq:full_asymptotic} by averaging
Eq.~\eqref{eq:EmAbar} over the number $m$ of relevant nodes with
distribution $P_N(m)$ (Eq.~\eqref{eq:p_N_m}).

Applying the inequalities
$\exp(-2x)\leq1-x$ for $0\leq x\leq1/2$ and
$1-x\leq\exp(-x)$ factorwise to Eq.~\eqref{eq:p_N_m} and using
$\sum_{j=1}^{m-1}j=m(m-1)/2$, we obtain
\begin{equation}
\frac{m}{N}\exp\left[-\frac{m^2}{N}\right]
\leq P_N(m)\leq
\frac{m}{N}\exp\left[-\frac{m^2}{4N}\right],
\label{eq:PN_bounds_end}
\end{equation}
where the lower bound holds for $m\leq N/2$ and the upper bound for
$m\geq2$.

Fix $0<\epsilon<1/2$. By Eq.~\eqref{eq:EmAbar}, there exists
$m_\epsilon$ such that for all $m\geq m_\epsilon$,
\begin{equation}
\exp\left[m^{1/2-\epsilon}\right]
\leq
\mathbb E_m[\bar A]
\leq
\exp\left[m^{1/2+\epsilon}\right].
\label{eq:EmA_epsilon_bounds}
\end{equation}
For the upper bound, the finitely many terms with $m<m_\epsilon$
contribute only $O(1)$, so Eqs.~\eqref{eq:PN_bounds_end}
and~\eqref{eq:EmA_epsilon_bounds} give
\begin{equation}
\mathbb E_N[\bar A]
\leq
O(1)+
\sum_{m=m_\epsilon}^{N}\frac{m}{N}
\exp\left[-\frac{m^2}{4N}+m^{1/2+\epsilon}\right].
\label{eq:EN_upper_sum}
\end{equation}
The exponent contains two competing terms. 
Setting $m\sim N^{\beta_+}$,
they scale as $N^{2\beta_+-1}$ and
$N^{\beta_+(1/2+\epsilon)}$, respectively.
Balancing these powers gives $\beta_+=2/(3-2\epsilon)$. 
At this saddle-point scale, both terms are of order $N^{\alpha_+}$, where
$\alpha_+ = ({1+2\epsilon})/({3-2\epsilon})$ and the exponent achieves its maximum.
Writing $m=N^{\beta_+}x$, the exponent becomes
\begin{equation}
-\frac{m^2}{4N}+m^{1/2+\epsilon}
=
N^{\alpha_+}
\left(-\frac{x^2}{4}+x^{1/2+\epsilon}\right).
\label{eq:upper_rescaling}
\end{equation}
The function in parentheses has a finite maximum $C_\epsilon$.
Since $m/N\leq1$ and there are at most $N$ terms,
\begin{equation}
\mathbb E_N[\bar A]
\leq
O(1)+N\exp\left[C_\epsilon N^{\alpha_+}\right]
=
\exp\left[
N^{\frac{1+2\epsilon}{3-2\epsilon}+o(1)}
\right].
\label{eq:EN_upper_bound}
\end{equation}

For the lower bound, since all terms in the ensemble average are nonnegative, Eqs.~\eqref{eq:PN_bounds_end}
and~\eqref{eq:EmA_epsilon_bounds} give, for any
$m_\epsilon\leq m\leq N/2$,
\begin{equation}
\mathbb E_N[\bar A]
\geq
\frac{m}{N}
\exp\left[-\frac{m^2}{N}+m^{1/2-\epsilon}\right].
\label{eq:EN_lower_single}
\end{equation}
Analogously to the upper bound, balancing the powers yields $\beta_-=2/(3+2\epsilon)$. Let $m = \lfloor xN^{\beta_-}\rfloor$ for any $x\in (0,1)$, then for sufficiently large $N$, $m_{\epsilon} \leq m \leq N/2$, and Eq.~\eqref{eq:PN_bounds_end} gives
\begin{equation}
\mathbb E_N[\bar A]
\geq
\frac{m}{N}
\exp\left[
N^{\alpha_-}
\left(-x^2+x^{1/2-\epsilon}+o(1)\right)
\right],
\label{eq:EN_lower_scaled}
\end{equation}
where $\alpha_-=(1-2\epsilon)/(3+2\epsilon)$.
Since $-x^2+x^{1/2-\epsilon}>0$ and $m/N$ is only polynomial in $N$,
\begin{equation}
\mathbb E_N[\bar A]
\geq
\exp\left[
N^{\frac{1-2\epsilon}{3+2\epsilon}+o(1)}
\right].
\label{eq:EN_lower_bound}
\end{equation}

Letting $\epsilon\to0$ yields in Eqs~\ref{eq:EN_upper_bound}, and \ref{eq:EN_lower_bound}
\begin{equation}
\mathbb E_N[\bar A]
=
\exp\left[N^{1/3+o(1)}\right],
\label{eq:EN_A_final_end}
\end{equation}
as stated in Eq.~\eqref{eq:full_asymptotic}.
The corresponding saddle-point scale is $m=N^{2/3+o(1)}$:
although a typical network has only $m=O(\sqrt N)$ relevant nodes,
rare networks with parametrically more relevant nodes determine the
ensemble-mean scaling.

\bibliographystyle{unsrt}

\bibliography{references}

\end{document}